\documentclass[conference,usenatbib]{basi}
\usepackage[T1]{fontenc}
\usepackage[british]{babel}
\usepackage[varg]{txfonts}
\usepackage{rotating}
\usepackage{dcolumn}
\begin{document}
\title{Estimation of mass of MAXI J1659-152 during its very first outburst with TCAF fits}
\author[Molla et~al.]%
       {Aslam Ali Molla$^1$\thanks{email: \texttt {aslam@csp.res.in}}, S.K. Chakrabarti$^{2,1}$, 
D. Debnath$^1$, S. Mondal$^1$, A. Jana$^1$, and  D. Chatterjee$^1$\\
       $^1$Indian Center for Space Physics, 43 Chalantika, Garia St. Rd., Kolkata, 700084, India.\\
       $^2$S. N. Bose National Centre for Basic Sciences, Salt Lake, Kolkata, 700098, India.}

\pubyear{2015}
\volume{12}
\pagerange{119-120}


\maketitle
\label{firstpage}

\begin{abstract}
We determine mass of MAXI J1659-152 using two different methods, 
viz:\textit{constant Normalization parameter point of view while fitting with TCAF}, 
\textit{$M_{BH}$-$\chi^2_{red}$} \textit{variation from TCAF fits}. 
We find a mass range of $4.06-7.93~M_\odot$ with a most probable value at $6 \pm 2 M_\odot$.
\end{abstract}



\vskip 0.1cm
\noindent{\bf 1. Introduction}
\vskip 0.05cm

MAXI J1659-152 was first observed by MAXI/GSC instrument on 
24th Sept. 2010 at the sky location of R.A. $= 16^h59^m10^s$, 
Dec $= -15^\circ 16'05''$ (Negoro et al., 2010). It showed its 
very first outburst in 2010, other than low-level activity in 2011 
which continued for $\sim 9$~months. The most acceptable 
ranges of distance and disk inclination angle are
$5.3-8.6$~kpc and $60-80$~deg. (Yamaoka et al. 2012 \& 
Kuulkers et al. 2013) respectively. Mass of the black hole candidate is 
estimated to be in the range $3.6-8~M_\odot$ (Yamaoka et al. 2012), 
$2.2-3.1~M_\odot$ (Kenna et al. 2011), $20\pm3~M_\odot$ (Shaposhnikov et al. 2011). 


\vskip 0.1cm
\noindent{\bf 2. Results and Concluding remarks}
\vskip 0.05cm

In Fig. 1a we see that the value of normalization parameter 
(61.94-86.71) remains at about a constant value (=70) throughout the outburst
as it should be. During the fitting we tried to fit data with acceptable reduced $\chi^2$
keeping Normalization roughly constant. This leads to a narrow range of possible mass
of the black hole. During fittng if we keep our all parameters to be free and maintain 
the normalization parameter to be constant, we find mass of black hole 
in the range of 5.14-7.93$M_\odot$.
In the second method, we fit 30 observation Ids with TCAF (Chakrabarti \& Titarchuk., 1995) 
model after the inclusion in XSPEC (Debnath et. al. 2014). 
During fitting we keep all the parameter free. After achieving the 
best fit (for details, see, Debnath et. al. 2015) we freeze all 
the parameter except the mass of black hole and tried to find 
the range of mass within which, value of reduced $\chi^2$ is 
acceptable (<2). In Fig. 1(b-c) we see that value of 
reduced $\chi^2$ behaves well within a mass range of 
4.06-7.87$M_\odot$. So, from variation of \textit{$M_{BH}$-$\chi^2_{red}$} 
behaviour we can bracket the mass of black hole within mass range of 4.06-7.87$M_\odot$.
Combining the results of these two different types of
estimates we obtain a single set of bounds for the mass. The most probable mass of 
black hole candidate MAXI J1659-152 appears to be $6\pm2~M_\odot$.

\begin{figure}
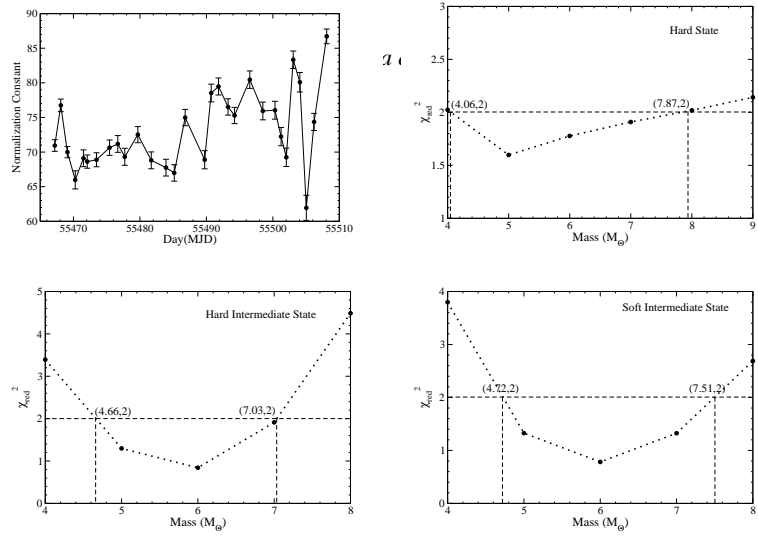

\vskip -1.5 cm
\centerline{\includegraphics[scale=0.8,width=4.5truecm]{Plot_norm.eps} \qquad
            \includegraphics[scale=0.8,width=4.5truecm]{Hard.eps}}
\vskip 0.5cm
\centerline{\includegraphics[scale=0.8,width=4.5truecm]{Hard_Intermediate.eps} \qquad
            \includegraphics[scale=0.8,width=4.5truecm]{Soft_Intermediate.eps}}
\vskip -0.3cm
\caption{ (a) Variation of normalization parameter with day, which is roughly constant (=70) over the entire outburst. In figure b, c and d the Variation of reduced $\chi^2$ with mass of black hole ($M_{BH}$) in solar mass ($M_\odot$) unit for different spectral states are shown, value of reduced $\chi^2$ remains good in between mass range 4.06$M_\odot$-7.87$M_\odot$ beyond which value of reduced $\chi^2$ is >2.}
\label{fig1}
\end{figure}

\end{document}